\def\year{2021}\relax
\documentclass[letterpaper]{article} 
\usepackage{times}  
\usepackage{helvet} 
\usepackage{courier}  
\usepackage[hyphens]{url}  
\usepackage{graphicx} 
\usepackage{natbib}  
\usepackage{caption} 
\usepackage[a4paper, margin=0.75in]{geometry}

\usepackage{color} 

\usepackage{multirow}
\usepackage{array}
\usepackage{subcaption}
\usepackage{colortbl}
\usepackage{amsmath}

\author {
    \normalsize Soham Poddar$^1$*, Mainack Mondal$^1$, Janardan Misra$^2$, Niloy Ganguly$^{1,3}$, Saptarshi Ghosh$^1$\\
    \textit{\small $^1$Department of Computer Science and Engineering, Indian Institute of Technology, Kharagpur, India}\\
    \textit{\small $^2$Accenture Labs, Bangalore, India}\\
    \textit{\small $^3$Leibniz University Hannover, Germany}
}

\title{Winds of Change: Impact of COVID-19 on\\  Vaccine-related Opinions of Twitter users}

\date{}

\newcommand{\reviewerchange}[1]{#1}
\newcommand{\new}[1]{#1}

\newcommand\blfootnote[1]{%
  \begingroup
  \renewcommand\thefootnote{}\footnote{#1}%
  \addtocounter{footnote}{-1}%
  \endgroup
}

\begin{document}
\maketitle
\thispagestyle{empty}

\vspace{-6mm}
\begin{abstract}

Administering COVID-19 vaccines at a societal scale has been deemed as the most appropriate way to defend against the COVID-19 pandemic. 
This global vaccination drive naturally fueled a possibility of Pro-Vaxxers and Anti-Vaxxers strongly expressing their supports and concerns regarding the vaccines on social media platforms. Understanding this online discourse is crucial for policy makers. This understanding is likely to impact the success of vaccination drives and might even impact the final outcome of our fight against the pandemic. 
The goal of this work is to improve this understanding using the lens of Twitter-discourse data. 
We first develop a classifier that categorizes users according to their vaccine-related stance with high precision (97\%).
Using this method we detect and investigate specific user-groups who posted about vaccines in pre-COVID and COVID times. 
Specifically, we identify distinct topics that these users talk about, and investigate how  vaccine-related discourse has changed between pre-COVID times and COVID times. 
Finally, for the first time, we investigate the change of vaccine-related stances in Twitter users and shed light on potential reasons for such changes in stance. 
Our dataset and classifier are available at \textcolor{blue}{\url{https://github.com/sohampoddar26/covid-vax-stance}} .

\end{abstract}
\blfootnote{\textbf{This work has been accepted to appear at International Conference on Web and Social Media (AAAI ICWSM) 2022.}}
\blfootnote{*\textit{Corresponding Author, email:} \url{sohampoddar@kgpian.iitkgp.ac.in}}

\vspace{-4mm}
\section{Introduction}\label{sec:intro}

\noindent 
The ongoing COVID-19 pandemic has affected 225M people till now (September 2021) and caused more than 4.6M deaths.
Fortunately, since December 2020 / January 2021, multiple pharmaceutical companies have put forward vaccines (e.g., AstraZenca, Pfizer, Moderna to name a few) that
are claimed to reduce the chance of COVID infection and fatality.\footnote{\url{https://tinyurl.com/WHO-vaccine-advice}}
Naturally, governments across the world are procuring and administering these vaccines to their citizens.

However, administering COVID-19 vaccines also has a key societal angle. To make these vaccines effective against COVID-19 at-scale and to eradicate the disease from a society, almost everyone in that society needs to consent to be vaccinated. In other words, only with near societal-scale vaccination, can we  achieve the ``herd immunity'' and eradicate the disease (and prevent countless deaths)\footnote{\url{https://tinyurl.com/WHO-COVID-immunity}}. 
These vaccines are naturally welcomed by many people. However, some have also speculated that the development of these vaccines might be rushed and  propelled by monetary and political reasons, thus indicating their \textit{hesitancy} towards COVID-19 vaccines~\citep{chou2020considering}. This hesitancy persists in large sections of the society in spite of Governments giving various incentives towards vaccination.

There exists a long-drawn debate about vaccines
since long before the onset of the COVID-19 pandemic.
The debate is between \textit{Anti-Vaxxers} who believe that vaccines do more harm than good, and \textit{Pro-Vaxxers} who support and promote the benefits of vaccines. Both the groups are known to actively propagate their views over social media~\citep{mitra2016understanding,gunaratne2019temporal}. 
It is important to know what the present attitude of the Anti-Vaxxer and Pro-Vaxxer communities is towards COVID-19 vaccines, and whether there has been any change in these attitudes in the pandemic times.
While there have been a few surveys and manual analyses that have attempted to answer these questions~\cite{funk2020intent,bonnevie2020quantifying}, to our knowledge, there has not been any prior attempt to devise automatic methods of answering these questions at scale using data from social media. 
It is important for policy makers to get answers to these questions at multiple stages of the pandemic, e.g., to figure out what the main concerns about vaccines are at different stages, so that they can take intervening steps accordingly. 
To our knowledge, this study is the first that (i)~devises  automated methods over data collected from social media (Twitter) to answer such questions, and (ii)~conducts a long-term (over 2 years) analysis of people's attitudes towards vaccines, both individually and collectively.
Specifically, we investigate two key questions --
(1)~What are user-attitudes towards COVID-vaccine as expressed by Anti-Vaxxers and Pro-Vaxxers on Twitter? Especially, are the anti-vax opinions about COVID-19 vaccines similar to the traditional anti-vax arguments that were prevalent in pre-COVID times?, and
(2)~Has the vaccine-related stance of some users changed over time due to the pandemic and why?

We make three important contributions in this work. 
First, we develop a dataset of COVID-vaccine related tweets labelled according to their vaccine stances, as detailed in Section~\ref{sec:classifier}.
Training a classifier on these tweets, we created a method to {\it categorize Twitter users according to their vaccine-related stance} as expressed through their tweets both in pre-COVID and COVID times, with 97\% precision.
We use this method to detect and investigate specific user-groups, e.g., Anti-Vaxxers and Pro-Vaxxers. 

Second, we use topic modeling 
to identify $12$ COVID-vaccine related topics/themes from the tweets of Pro-Vaxxers and Anti-Vaxxers. 
We further analyze the relative presence of these topics in vaccine-related online discourse in
different time periods -- pre-COVID, COVID and COVID-vax time periods (described in Section~\ref{sec:data_collection}). 
Through this analysis, we find that some anti-vax themes such as {\it the vaccine development being rushed} and {\it concerns about vaccines being ineffective} have become more frequently discussed in COVID times (compared to the pre-COVID times).

Third, we show that there exists a set of users who have changed their original vaccine-related stance in the COVID times (e.g., from Anti-Vaxxers to Pro-Vaxxers, or vice-versa).
Understanding this set of users, whom we call {\bf stance-changed users}, can be important for the authorities who are attempting to maximize the coverage of vaccines.
Though some prior works have shown that people have changed their stances towards COVID-19 vaccines~\cite{funk2020intent,bonnevie2020quantifying}, no prior work has systematically explored such users and {\it why} they changed their stance.
In this work, we characterize different sub-groups of stance-changed users.
Furthermore, we explore {\it possible reasons of such stance-change}. To this end, we uncover correlations between the stance-change of a user and life-events (e.g., loss of a close relative) as well as change in stance of the users' social contacts.

We believe that the insights obtained from this study will help authorities better understand how vaccine-related stances of users are changing in pandemic times, and will help them to formulate effective policies to increase vaccine adoption by the masses. Our dataset and classifier are available at \textcolor{blue}{\url{https://github.com/sohampoddar26/covid-vax-stance}} .

\section{Related Works}\label{sec:rel_work}

The dichotomy of opinions on vaccines between \textit{Anti-vaxxers} and \textit{Pro-vaxxers} on social media has been identified in many prior works~\citep{mitra2016understanding,gunaratne2019temporal}. 
Both these groups have been seen to be quite strong in their stance,
each believing the other to indulge in conspiracies~\cite{yuan2019examining,cossard2020falling}. We briefly survey a few relevant works in this section.

\vspace{2mm}
\noindent {\bf Vaccine-stance detection methods:} 
Several studies have attempted to classify social media posts (e.g., tweets) based on their stance towards vaccines, using  hashtag counts~\cite{gunaratne2019temporal}, machine learning models~\cite{mitra2016understanding, yuan2019examining} and neural  models~\cite{muller2020covid, cotfas2021longest}.
Some works have also provided labeled datasets for this classification, as detailed in Section~\ref{sec:classifier}.
Some studies have also tried to classify the stances of {\it users} based on percentages of posted tweets of different stances~\cite{mitra2016understanding} as well as by community detection algorithms~\cite{yuan2019examining, gunaratne2019temporal}.

\vspace{2mm}
\noindent \textbf{Social media discourse on COVID-19 vaccines:}
With the onset of the COVID-19 pandemic, people have naturally resorted to sharing their emotions on social media~\cite{lwin2020global}.
Tough times give rise to several conspiracy theories and the spread of misinformation about the disease has risen~\cite{cinelli2020covid}. 
The pandemic has also led to decline of public trust on the healthcare system and a corresponding huge increase in hesitancy towards COVID-19 vaccines~\cite{johnson2020online,bonnevie2020quantifying}.

A few studies have tried to understand which demographics of people are more likely to be hesitant towards the vaccines~\cite{malik2020determinants, thunstrom2020hesitancy,funk2020intent}.
Some studies have also tried to understand the reasons behind the rise of this hesitancy~\cite{jamison2020not,bonnevie2020quantifying,praveen2021analyzing}, some major reasons being distrust towards pharmaceutical industries and the potential side effects~\footnote{\url{https://tinyurl.com/global-attitudes-covid-vaccine}} (which we also observe in this study, as detailed in later sections).

\vspace{2mm}
\noindent {\bf This work:}
Though some prior works showed that people have changed their stances towards COVID-19 vaccines, these works relied on surveys~\cite{funk2020intent} or manual coding~\cite{bonnevie2020quantifying}. 
To our knowledge, no automated methods have been developed for such analyses at scale. 
Also, it is unclear {\it why} different users have changed their stance, and what are the characteristics of such users.
This work is the first attempt to bridge this gap where we study how and why vaccine-related opinions of users have changed between pre-COVID times and COVID times.

\section{Data collection}
\label{sec:data_collection}

In this work, we planned to study tweets posted since Pre-COVID times. Since the Twitter API only allows collecting tweets posted in the last $\sim 7$ days,  we used the Twint library\footnote{\url{https://github.com/twintproject/twint}} to collect old tweets. The Twint library has been used by several prior works for collecting older tweets~\citep{dutta2021analyzing,punuru2020cultural,nuzhath2020covid}.
We used this library to collect a total of 15.7 million English tweets from 4 million distinct users over the duration January 2018 -- March 2021. 
We consider this duration divided into three stages, as described below.

\vspace{2mm}
\noindent \textbf{(1) Pre-COVID period (Jan 2018 to Dec 2019):} 
This period is before the COVID pandemic, when people talked about vaccines in general. To get a good coverage of vaccine-related tweets, we used some generic keywords such as `vaccine', `vaccination' `vaxxer', etc. along with an extensive set of Pro-vax and Anti-vax hashtags provided by~\citet{gunaratne2019temporal}. 
Using these keywords, we collected 2.9M English tweets posted in this period.

\vspace{2mm}
\noindent \textbf{(2) COVID period (2020), and (3) COVID-vax period (Jan-Mar 2021):}
The \textit{COVID period} signifies the year 2020, when COVID-19 was declared as a pandemic and various organizations started working on developing vaccines. 
The \textit{COVID-vax period} signifies the first quarter of 2021 (January--March 2021) when COVID-19 vaccines started to be administered globally. 
We consider these two time-periods separately since people's perceptions about COVID-19 vaccines may have varied between these time periods (before and after a vaccine was available).

For these periods, in addition to the keywords used for Pre-COVID period, 
we also fetched tweets using some generic keywords such as `corona vaccine' and `covid vaccine' and the names of popular vaccines and their manufacturers, such as `astra zeneca', `pfizer', `moderna', etc.
We collected 6.6M English tweets from the COVID period (Jan--Dec 2020) and 6.2M English tweets from the COVID-vax period (Jan--Mar 2021). Evidently, the social media discourse on vaccines has increased manifold in 2021 after the administration of COVID vaccines has started. 

\vspace{2mm}
\noindent \textbf{Checking for bias in tweets collected by Twint:}

Since we did not use the official Twitter API to collect tweets, we check if our collected data contains any bias. 
To this end, we collected the tweets posted during June 2021 using both the Twitter API (fetched in real-time) and the Twint library (fetched in August 2021) using a common set of search-query terms. 
While the Twitter API collected 4.8M distinct tweets, the Twint library collected 745K distinct tweets during the same period. 
Thus, Twint collects 15.5\% of all tweets collected by the Twitter API. 
Note that, Twint does not collect retweets; so we have not considered the retweets collected by the Twitter API. 
Next, we compared the set of tweets collected by the two methods as follow:

\vspace{2mm}
\noindent $\bullet$ \textbf{Representativeness w.r.t. time:} On each of our observed days, the fraction of tweets collected by the Twint library is relatively uniform (close to 15\% of what is collected via Twitter API). This gives us confidence that the set of tweets collected by the Twint library is temporally representative.

\vspace{2mm}
\noindent $\bullet$ \textbf{Representativeness w.r.t. tweet popularity:} Next we check if the Twint library is biased towards more popular tweets. We measure the popularity of a tweet by its {\it retweet-count} that is stated in the tweet object. 
Among the tweets collected by Twitter API, 93.0\% had less than 10 retweets and 99.1\% had less than 100 retweets. 
In comparison, among the tweets collected by Twint, 90.4\% had less than 10 retweets and 98.8\% had less than 100 retweets.
The numbers are quite similar for both Twitter API and Twint. Thus, the set of tweets collected by Twint does not have bias towards more popular tweets.

Thus, though we could fetch only 15\% of all the posted tweets, the collected tweets do not have much apparent biases.
Also we believe that the stance-detection methods used in this work (described in the next section) are robust towards small variations in data w.r.t time or tweet
popularity, due to the conservative thresholds we use in the classifiers. 

Since the Twitter API does not allow access to tweets older than $\sim 7$ days\footnote{\url{https://developer.twitter.com/en/docs/twitter-api/v1/tweets/search/overview}}, we needed to use some library like Twint to collect older tweets, as has been done by several prior longitudinal studies~\citep{dutta2021analyzing,punuru2020cultural,nuzhath2020covid}. 
Also, many prior studies have been based on the 10\% and even the 1\% Twitter random sample (provided by Twitter). Hence, we believe that the mostly unbiased 15\% tweet sample (that Twint gave us) is sufficient and appropriate for the present study.

\vspace{-2mm}
\section{Identifying users' stance towards vaccines}\label{sec:classifier}
\vspace{-1mm}

This section discusses our approach to classify users based on their stance towards vaccines during the three time periods (pre-COVID, COVID, and COVID-vax). 
Since a user's stance may have changed over time, we do {\it not} attempt to do this classification based on the profile attributes of the user. 
Instead, we first develop a tweet-classifier to predict the stance of individual tweets (Section~\ref{sub:tweet_classficiation}), and then use this classifier to identify a user's stance during a certain time period from the tweets posted during that period (Section~\ref{sub:user_classification}).

\vspace{-2mm}
\subsection{Developing a tweet classifier}\label{sub:tweet_classficiation}

\noindent {\bf Annotating tweets and other datasets:}
There exist a few publicly available datasets that have labelled tweets as \textit{Anti-Vax}, \textit{Pro-Vax} or \textit{Neutral}  according to their stance towards vaccines. 
\citet{muller2019crowdbreaks} provided a large set of tweets about vaccines before the onset of the COVID-19 pandemic, while \citet{cotfas2021longest} provided a dataset of tweets about COVID-19 vaccines collected during a month between November and December of 2020.
We call these two datasets the {\bf Muller} and {\bf Cotfas} datasets respectively.\footnote{Both these datasets only provide the tweet-ids, so we could only work with those tweets in the datasets that we could re-crawl in 2021 (i.e., have not been deleted earlier).}

Since the Cotfast dataset only consists of tweets from one particular month, it might not cover different types of tweets posted over the entire discourse about COVID-19 vaccines. Hence we wanted to extend this dataset with tweets from a broader timeline. To this end, we randomly selected $1,700$ tweets posted between January 2020 and March 2021, to be annotated into the three classes.
For the annotation, we employed three annotators on the crowdsourcing platform Prolific (\url{https://www.prolific.co/}). To ensure reliable annotation, we employed only annotators who are proficient in English and
have participated in more than $1000$ tasks with 100\% acceptance rate on the platform (top 1\% annotators).

\begin{table}[t]
	\centering
    \small
	\begin{tabular}{|m{45mm}|c|c|c|c|}
		\hline
		\textbf{Dataset} & \textbf{\#tweets} & \textbf{Anti-Vax} & \textbf{Pro-Vax} & \textbf{Neutral}\\
		\hline
		Muller & 18.5k & 10.5\% & 40.3\% & 49.3\%\\
		\hline
		Cotfas & 2,792 & 28.3\% & 35.5\% & 36.2\%\\
		\hline
		Cotfas + our dataset (majority) & 4,398 & 24.6\% & 38.2\% & 37.2\%\\
		\hline    
		Cotfas + our dataset (unanimous) & 3,577 & 26.0\% & 38.7\% & 35.3\%\\
		\hline    
	\end{tabular}
	\vspace{-2mm}
	\caption{Summary of different datasets used for training and testing of the classifiers, each of them having 3 class labels}
	\vspace{-2mm}
	\label{tab:datasets}
\end{table}

Among the 1700 annotated tweets, $94$ tweets were labelled different by all three annotators, $785$ tweets had unanimous agreement, and $821$ tweets were labelled the same by two annotators and different by the other. 
\reviewerchange{
We discarded the tweets where all three annotators disagreed. We create two datasets by merging our annotated tweets with the tweets in the Cotfas dataset --
in one dataset, we added the 1,606 tweets having majority agreement to the Cotfas dataset;
in the other dataset, we add the 785 tweets having unanimous agreement to the Cotfas dataset.
Thus we have four datasets summarized in Table~\ref{tab:datasets}.
}

\begin{table}[t]
	\centering
\small
	\begin{tabular}{|l|c||c||c||c|}
		\hline
		& \multicolumn{1}{c||}{Muller} & \multicolumn{1}{c||}{Cotfas} &  \multicolumn{1}{c||}{\shortstack{Cotfas + our\\ dataset  (majority)}} & \multicolumn{1}{c|}{\shortstack{Cotfas + our \\dataset (unanimous)}}\\
		\hline
		\hline    
		SVM & 0.628 & 0.751 & 0.642 & 0.695 \\
		\hline
		FastText & 0.650 & 0.735 & 0.624 & 0.683 \\
		\hline
		BERT-large & 0.661 & 0.763 & 0.670 & 0.720 \\
		\hline
		CT-BERT & \textbf{0.687} & 0.768 & 0.744 & 0.797 \\
		\hline \hline
		CT-BERT++ & - & \textbf{0.858} & \textbf{0.775} & \textbf{0.825} \\
		\hline    
	\end{tabular}
	\vspace*{-2mm}	
	\caption{Average macro F1-score over stratified 5-fold cross-validation, for the classifiers on the four datasets. ``CT-BERT++'' is CT-BERT finetuned on the dataset by Muller et al, before cross-validation on the other datasets. The best scores are marked in bold.}
	\vspace*{-4mm}
	\label{tab:classification_scores}
\end{table}

\begin{table}[t]
	\centering
    \small
	\begin{tabular}{|c|l|c|c|c|}
		\hline
		& & \multicolumn{3}{c|}{Trained on}\\
		\cline{3-5}
		& & Muller & Cotfas & Our dataset\\
		\hline
		\multirow{3}{*}{\rotatebox[origin=c]{90}{Tested on}} & Muller & - & 0.430 & 0.471 \\
		\cline{2-5}
		 & Cotfas & 0.723 & - & 0.347 \\
		\cline{2-5}
		 & Our dataset & 0.583 & 0.475 & - \\
		\hline    
	\end{tabular}
	\vspace*{-2mm}
	\caption{\reviewerchange{Average macro F1-score of the CT-BERT classifier trained and tested on different datasets.}}
	\vspace*{-2mm}
	\label{tab:cross_testing}
\end{table}

\vspace{2mm}
\noindent {\bf Classifier models:} To classify tweets into the three classes, we used a variety of classification models, including Support Vector Machines (SVM) with TF-IDF feature vectors of tweets,  
the supervised FastText classifier~\cite{joulin2016bag} along with some state-of-the-art BERT-based models. 
Specifically, we tried out the standard `bert-large'~\cite{devlin2018bert},
and the `covid-twitter-bert-v2'~\citep{muller2020covid} model which has been pre-trained specifically on COVID-related tweets (which we refer to as CT-BERT).\footnote{We also tried some other models including Random Forest and `bertweet-covid19-base-uncased'~\cite{bertweet}. These models performed worse than the reported ones in comparison; hence omitted for brevity.}
We used the implementations of these models provided by the HuggingFace library~\citep{wolf-etal-2020-transformers}.
Each of the BERT-based models were paired with a classification head to perform three-class classification.

\vspace{2mm}
\noindent {\bf Evaluation of tweet classification:}
We performed stratified 5-fold cross validation on the four datasets, and evaluated the classifiers using the average macro-F1 scores across the 5-folds.
Table~\ref{tab:classification_scores} (first four rows) presents the cross validation performance of the classifiers. On each of the four datasets, `covid-twitter-bert-v2'~(CT-BERT) performs the best.

We also performed cross-dataset testing, where we trained the CT-BERT model on one dataset and tested on other datasets.
Table~\ref{tab:cross_testing} shows the macro-F1 scores of CT-BERT in cross-dataset setting. 
The model does not perform too well when trained and tested on different datasets, possibly because the time-scale of the datasets are very different -- the Muller dataset is of pre-Covid times, Cotfas dataset contains only tweets posted in November 2020, while our dataset spans January 2020 through March 2021 -- and hence the nature of the tweets is somewhat different.

\vspace{2mm}
\noindent {\bf Final classifiers for the whole collected data:}
We intend to develop an accurate classifier for vaccine-related tweets posted over a long time span (January 2018 through March 2021). 
Since we saw that different datasets have different characteristics (as discussed above), we decided to use data from all the different datasets to design a more accurate classification scheme.


Since the Muller dataset is an order of magnitude larger than the other two datasets, we try a variant of the CT-BERT model that is initially finetuned on the Muller dataset.
We refer to this model as `CT-BERT++' (with weights updated via finetuning over the Muller dataset). 
We evaluate the CT-BERT++ model over the other datasets (apart from Muller) via stratified 5-fold cross validation, whose results are shown in the last row of Table~\ref{tab:classification_scores}. 
The CT-BERT++ model performs better than CT-BERT (and all other classifiers) over all the COVID-specific datasets, achieving a macro-F1 score of $0.825$ on the Cotfas dataset extended with the unanimously labelled tweets in our dataset. 
On the Cotfas dataset extended with our tweets having majority agreement, our model had a average macro-F1 score of $0.775$, which is in line with the original work~\citep{cotfas2021longest}.

Based on the results in Table~\ref{tab:classification_scores}, we decided to use two different classifiers to classify tweets from the pre-Covid times and the Covid-times.
To predict stance of tweets from the \textit{pre-COVID period}, we use the CT-BERT classifier fine-tuned over the Muller dataset (which achieved the best Macro F1-score over the Muller dataset in Table~\ref{tab:classification_scores}). 
To predict the stance of tweets from the \textit{COVID period} and the \textit{COVID-vax period}, we use CT-BERT++ finetuned again over the Cotfas dataset extended with the $785$ unanimously labelled tweets in our dataset.\footnote{We manually checked a sample of the tweets in our dataset, that had majority but not unanimous agreement. We observed that the stance of many such tweets were not very clear and could be open to interpretation; hence we preferred to use only the tweets having a unanimously agreed label for training the final classifier.} 
Using these classifiers we predicted the stance of all the tweets we had collected (described in Section~\ref{sec:data_collection}). The distribution of stances is given in Table~\ref{tab:tweet_stance_distribution}.

\vspace{2mm}
\noindent {\bf Checking for classifier bias:}
We check for any potential biases that the final classifier may have with respect to certain specific words in the data. 
To this end, we followed the {\it SPCPD bias word detection strategy}~\cite{badjatiya2019stereotypical} -- 
we ran the final classifier (CT-BERT++) on each individual {\it word} from the datasets, and checked which words are classified as Anti-Vax or Pro-Vax. 
Out of a total of 37,127 distinct words in all the datasets combined, the classifier does {\it not} label Anti-vax/Pro-vax for as many as 37,011 words, indicating {\it no bias with respect to 99.7\% of the words}. 
Only $116$ words were predicted as Anti-Vax or Pro-Vax. These words were  manually observed by a human annotator. All these words are found to be hashtags that are used almost exclusively with the corresponding stances.
For example, hashtags such as `\#refusePoisonVaccines', `\#dontTrustVaccines', and `\#vaccinesKill'
were labeled as Anti-Vax, 
whereas hashtags such as `\#getVaccinated', `\#vaccinesSaveLives', and `\#vaccinateToStopTheSpread' were labeled as Pro-Vax. 
We believe these keywords are indeed strongly associated with the corresponding stances.
Thus we did not find any unjustified bias in the classifier.
Also, we ran the classifier on the datasets after removing these hashtags, and the performances remained almost the same (e.g., on the [Cotfas + our unanimously labeled tweets] dataset, there was a decrease of only 0.003 in the macro-F1 score).

\begin{table}[t]
	\centering
    \small
	\begin{tabular}{|l|c|c|c|}
		\hline
		\textbf{Time Period} & \textbf{\#tweets} & \textbf{Anti-Vax} & \textbf{Pro-Vax}\\
		\hline
		\hline
		Pre-COVID (2018-19) & 2.9M & 19.9\% & 47.0\%\\
		\hline
		COVID (2020) & 6.5M & 20.0\% & 27.6\% \\
		\hline
		COVID-vax (Jan-Mar 2021) & 6.3M & 17.1\% & 41.7\%\\
		\hline
		Overall & 15.7M & 18.8\% & 36.9\% \\
		\hline
	\end{tabular}
	\vspace*{-2mm}
	\caption{\reviewerchange{Distribution of predicted stances of all our collected tweets during the different time periods.}}
	\label{tab:tweet_stance_distribution}
	\vspace*{-4mm}
\end{table}

\subsection{Identifying vaccine-related leaning of users}
\label{sub:user_classification}

\noindent \textbf{Identifying Pro-Vaxxers and Anti-Vaxxers:} 
We use the tweet-classifiers described above to identify the stance of every tweet posted by a user. 
We then label a user as \textit{Anti-Vaxxer} or \textit{Pro-Vaxxer} during a specific time period if they posted at least 3 vaccine-related tweets during that period, out of which more than $\tau$ fraction is labeled as \textit{Anti-Vax} or \textit{Pro-vax} respectively. 
We filter out users who posted less than 3 tweets as it is difficult to accurately estimate their stance from so few tweets.
We use $\tau = 0.7$ similar to~\cite{mitra2016understanding} to get good precision in detecting user-stances. 
Using this method, we identified several users as \textit{Anti-Vaxxers} and \textit{Pro-Vaxxers} during the three time periods, as summarised in Table~\ref{tab:users_identified}.

\begin{table}[t]
	\centering
    \small
	\begin{tabular}{|l|c|c|c|}
		\hline
		\textbf{Time Period} & \textbf{Anti-Vaxxers} & \textbf{Pro-Vaxxers} & \textbf{Unidentified}\\
		\hline
		Pre-COVID (2018-19) & 14,189 & 57,905 & 91,800\\
		\hline
		COVID (2020) & 46,872 & 54,771 & 362,765\\
		\hline
		COVID-vax (2021) & 36,768 & 124,053 & 304,475\\
		\hline
	\end{tabular}
	\vspace*{-2mm}
	\caption{\reviewerchange{Number of users identified as Pro-Vaxxers and Anti-Vaxxers during the different time periods. Last column states the number of users who posted 3 or more tweets during the corresponding time-period, but their stance could not be classified due to the classifier thresholds.}}
	\label{tab:users_identified}
	\vspace*{-2mm}
\end{table}

\begin{table}[t]
	\centering
    \small
	\begin{tabular}{|l|c|c|}
		\hline
		\textbf{Time Period} & \textbf{Anti-Vaxxers} & \textbf{Pro-Vaxxers} \\
		\hline
		Pre-COVID (2018-19) & 50/50 & 50/50 \\
		\hline
		COVID (2020) & 50/50 & 48/50 \\
		\hline
		COVID-vax (2021) & 44/50 & 49/50 \\
		\hline
	\end{tabular}
	\vspace*{-2mm}	
	\caption{Evaluation of user classification: number of correctly classified users (according to human annotators) out of $50$ randomly sampled users from each group as identified by the classifier. Overall precision = 97\% over the $300$ users checked.}
	\vspace{-4mm}
	\label{tab:user_classification_eval}
\end{table}

\vspace{1mm}
\noindent \textbf{Evaluation of user classification:} To evaluate the user classification, we randomly sampled $50$ users from each of \textit{Anti-Vaxxer} and \textit{Pro-Vaxxer} groups, as predicted by the classifier for each of the three time periods. 
Thus, overall, we selected $300$ users for this evaluation. 
Two annotators proficient in English and use of Twitter checked all the tweets posted by these users during the corresponding time periods, along with the users' Twitter profiles to determine their actual stance towards vaccines. 

Table~\ref{tab:user_classification_eval} shows the results of the user-classification evaluation.
Both annotators agreed that all the $100$ users from the pre-COVID period were classified correctly. 
For the COVID period, $98$ were correct, the other two users being ambiguous (i.e., at least one of the annotators was unsure of the user's actual stance). 
For the COVID-vax period, $93$ out of the $100$ users checked were correctly classified, while $6$ Pro-Vaxxers were wrongly classified as Anti-Vaxxers (both annotators agreed in all cases). 
This relatively higher misclassification rate for the COVID-vax period is because during this period, many users criticize the government for mismanagement of vaccination programmes, and 
the classifier wrongly classified some of these users as Anti-Vaxxers.

\new{Overall our method of user classification correctly identified the vaccine-related stance of $291$ out of $300$ users, giving a precision of 97\%.} 
We believe this performance is sufficiently good for the present study.

\vspace{1mm}
\noindent \textbf{Change in numbers of Anti-Vaxxers and Pro-Vaxxers:}
\reviewerchange{
Table~\ref{tab:users_identified} shows the variations in the number of Anti-Vaxxers and Pro-Vaxxers over the three time periods, while Table~\ref{tab:tweet_stance_distribution} shows the variations in distribution of Anti-Vax and Pro-Vax tweets. 
From these tables, it is evident that a lot more users have started discussing about vaccines since 2020, than in the pre-COVID times.}
Table~\ref{tab:users_identified} shows that there was a substantial increase in the number of Anti-Vaxxers in the COVID period (2020), owing to the uncertainty and delay in vaccination during that period.
However, once vaccination started in the COVID-vax period, a much larger number of users are seen to post Pro-vax opinions 
(even after the misclassfication of some Pro-Vaxxers as Anti-Vaxxers in the COVID-Vax period, as stated in Table~\ref{tab:user_classification_eval})

\new{
It can be noted that, in this paper, the set of `Anti-Vaxxers' may include a broader group of users than the traditional group studied in previous works prior to 2020. 
Anti-Vaxxers traditionally mean people who are against the intake of all vaccines in general. Whereas, in COVID times, people have shown different levels of hesitancy/unwillingness to take the COVID-19 vaccines (or even a particular COVID-19 vaccine) due to various reasons (as discussed in Section~\ref{sec:discourse}), even though they might not be against other vaccines. 
}

\vspace{2mm}
\noindent \textbf{Summary:} We designed a highly \reviewerchange{precise} user-classification method, using which we classified the vaccine-stance of tens of thousands of Twitter users, over the three time periods.

\section{Discourse regarding vaccines on Twitter}\label{sec:discourse}

\begin{table*}[!ht]
    \centering
    \footnotesize
    \begin{tabular}{|m{5.3cm}|m{5.3cm}|m{5.3cm}|}
        \hline
        \textbf{Pre-COVID period (2018-19)} & \textbf{COVID period (2020)} & \textbf{COVID-vax period (2021)} \\
        \hline \hline

        \rowcolor[gray]{0.8}
        \multicolumn{3}{|c|}{\textbf{Topic 1 [Health Concerns]:} Heath and safety issues -- Concerns about the ingredients and side effects of vaccines, including deaths caused.}\\
        \hline
        \rowcolor[gray]{0.9}
        \multicolumn{3}{|c|}{\textit{Top Words:} die, people, reaction, kill, receive, safe, effect, health, reaction, danger}\\
        \hline

        \textbf{[19.0\%]} Perhaps we now have the link between vaccination and autism spectrum disorder, the link being the inclusion of an aluminium adjuvant in the vaccine. &
        \textbf{[15.1\%]} I won’t take one derived from any aborted fetal tissues. Astra Zeneca does come from a cell line from aborted fetal tissues.  No way I’m taking that. & 
        \textbf{[18.3\%]} Herd immunity would serve us better and protecting the elderly and vulnerable. No need for a vaccine or experimental gene therapy \\
        \hline \hline

        \rowcolor[gray]{0.8}
        \multicolumn{3}{|c|}{\textbf{Topic 2 [Against Mandatory]:} The user is against mandatory vaccination -- talks about freedom and choice.}\\
        \hline
        \rowcolor[gray]{0.9}
        \multicolumn{3}{|c|}{\textit{Top Words:} want, force, mandatory, time, body, right, mandate, passport, refuse, court}\\
        \hline
        
        \textbf{[17.6\%]} If people WANT to get vaccinated, that's their choice. Anti-vaxxers are AGAINST gov't mandates and FOR informed consent and choice. & 
        \textbf{[12.1\%]} Biden is going to do his best to get a mandatory Covid vaccine  . I feel very bad for what's going to happen in America & 
        \textbf{[13.0\%]} The first time I'm turned away from anywhere for not having a Vaccine Passport. It's unconstitutional. My body my choice. \\
        \hline \hline

        \rowcolor[gray]{0.8}
        \multicolumn{3}{|c|}{\textbf{Topic 3 [Big Pharma]:} Dissatisfied with the role of pharmaceutical companies -- they are tricking people, trying to make money.}\\
        \hline
        \rowcolor[gray]{0.9}
        \multicolumn{3}{|c|}{\textit{Top Words:} pfizer, big, company, ask, pharma, immune, moderna, money, drug, manufacture}\\
        \hline

        \textbf{[17.4\%]} Big Pharma lines their pockets, doctors get bonuses for having 63\% of their patients vaccinated & 
        \textbf{[13.3\%]} Sure, let’s trust Bill Gates, a non-doctor billionaire who pays the media instead & 
        \textbf{[9.9\%]} If you suspect you have a slight breathing issue.... Run Away from astrazeneca vaccine... 
        \\
        \hline \hline

        \rowcolor[gray]{0.8}
        \multicolumn{3}{|c|}{\textbf{Topic 4 [Political]:} The vaccine is political in nature -- Governments are promoting their own agenda.}\\
        \hline
        \rowcolor[gray]{0.9}
        \multicolumn{3}{|c|}{\textit{Top Words:} trump, push, govern, lie, world, russian, gates, bill, fda, country}\\
        \hline
        
        \textbf{[9.8\%]} Vaccines is not ``allowed'' to be listed as cause of death. Genocide by the American Govt. & 
        \textbf{[12.2\%]} the CDC is spewing govt scripted propaganda and lies daily 
        & 
        \textbf{[10.8\%]} And no mention of opposing the terrible vaccine passports govt have plans for.\\
        \hline \hline

        \rowcolor[gray]{0.8}
        \multicolumn{3}{|c|}{\textbf{Topic 5 [Ineffective]:}  Vaccines don't work -- Skeptical about the effectiveness of vaccines}\\
        \hline
        \rowcolor[gray]{0.9}
        \multicolumn{3}{|c|}{\textit{Top Words:} work, will, get, stop, mask, lockdown, protect, effect, immune, mutate}\\
        \hline
        
        \textbf{[8.6\%]} MMR is a failure. Merck scientists are in court over lying about the efficacy of Mumps portion. &
        \textbf{[13.8\%]} researchers haven’t cured HIV in 40 years, so how could a vax for COVID be effective? &
        \textbf{[16.2\%]} I just read an article explaining that the first Covid-19 dose is only 52\% effective.
        \\
        \hline \hline

        \rowcolor[gray]{0.8}
        \multicolumn{3}{|c|}{\textbf{Topic 6 [Rushed]:} Vaccine development is being rushed -- Not enough testing has been done}\\
        \hline
        \rowcolor[gray]{0.9}
        \multicolumn{3}{|c|}{\textit{Top Words:} test, experiment, long, year, trial, medic, rush, risk, study, approve}\\
        \hline

        \textbf{[8.3\%]} This vaccine has NOT been tested for MUTAGENIC potential, for CARCINOGENIC potential, or for IMPAIRMENT of FERTILITY. &
        \textbf{[12.3\%]} What’s the rush? How many are actually dying here from Covid rather than another illness? Better to be safe than sorry with the politically timed vaccine &
        \textbf{[11.9\%]} Mark have you seen the side effects of the untested vaccine  Why would anyone be jumping or even in that queue Covid 19 has a 99.8\% survival rate \\
        \hline \hline 
        
        \rowcolor[gray]{0.8}
        \multicolumn{3}{|c|}{\textbf{Topic 7 [Shedding]:} Virus Shedding -- People who take vaccines will transmit the disease to others}\\
        \hline
        \rowcolor[gray]{0.9}
        \multicolumn{3}{|c|}{\textit{Top Words:} shed, people, flu, vax, live, mrna, fact, spread, cause, sick}\\
        \hline
         
        \textbf{[13.2\%]} where the measles come from, not the unvaccinated.. but the recently vaccinated  the mmr can shed for upto 21 days through recently vaccinated children. & 
        \textbf{[9.6\%]} do not take the vaccine. the vaccine will contain live virus, and people who get the vaccine, will shed the live virus and compound the problem. & 
        \textbf{[8.8\%]} we shouldn't be surprised if these new 'variant' strains are the vaccinated shedding the virus, or the virus mutating because of the vaccine. \\
        \hline\hline

        \rowcolor[gray]{0.8}
        \multicolumn{3}{|c|}{\textbf{Topic 8 [Deeper Conspiracy]:} Disease is a Hoax, Tracking Chips and talks of other conspiracies (other than just money making)}\\
        \hline
        \rowcolor[gray]{0.9}
        \multicolumn{3}{|c|}{\textit{Top Words:} need, people, risk, control, believe, old, vulnerable, death, population, mask}\\
        \hline
        
        \textbf{[6.0\%]} \#Vaccines are not for immunity. They are for POPULATION CONTROL aka \#eugenics. & 
        \textbf{[11.6\%]} COVID is about ``control'' \& bringing in ``Trojan horse'' vaccines from the loving Bill Gates &
        \textbf{[11.1\%]} get to the end game implant us with tracking chips and drug us to become the brainwashed robots \\
        \hline
    \end{tabular}
    \vspace{-2mm}
    \caption{Broad topics posted by Anti-Vaxxers, along with the percentage of tweets in that topic posted during a time period (within square brackets) and excerpts from sample tweets from the corresponding time periods.
    }
    \vspace{-4mm}
    \label{tab:annotation_themes_anti}
\end{table*}

In this section, we identify major themes/topics of discussion on vaccines across the different time-periods, by Anti-Vaxxers and Pro-Vaxxers.

\subsection{Method for identifying vaccine-related topics}

Given a particular set of tweets (e.g., the tweets posted by Anti-Vaxxers), we adopt the topic modeling-based method suggested in~\cite{bozarth2020higher}. 
The method involves four steps:
(1)~identifying candidate topics and topic-words through vanilla LDA, where we estimate the number of topics using topic coherence or perplexity scores~\cite{topic-coherence-newman}, 
(2)~manually mapping the identified topics and topic-words to a smaller subset of coherent topics and associated seed-words, 
(3)~using Labelled-LDA~\cite{ramage2009labeled} with these seed-words as input, and finally
(4)~manually examining the topics generated by Labelled LDA and combining topics that share the same theme and removing topics without coherence. 
This approach finally produced $8$ distinct {\it Anti-vax topics} from the tweets posted by Anti-Vaxxers and $4$ {\it Pro-vax topics} from the tweets of Pro-Vaxxers. 
\new{Overall, 80\% of the tweets posted by Pro-vaxxers and 83\% of tweets posted by Anti-vaxxers could be assigned a topic by the above method. 
We also computed the relative frequency of the topics in the three different time periods.}

\subsection{Opinions expressed by different user-groups}

\begin{table*}[!t]
    \centering
    \footnotesize
    \begin{tabular}{|m{5.3cm}|m{5.3cm}|m{5.3cm}|}
        \hline
        \textbf{Pre-COVID period (2018-19)} & \textbf{COVID period (2020)} & \textbf{COVID-vax period (2021)} \\
        \hline \hline

        \rowcolor[gray]{0.8}
        \multicolumn{3}{|c|}{\textbf{Topic 1 [Want Vaccines]:} User wants to get vaccinated or has got the vaccine.}\\
        \hline
        \rowcolor[gray]{0.9}
        \multicolumn{3}{|c|}{\textit{Top Words:} get, time, shot, feel, protect, thing, soon, receive, prevent, health}\\
        \hline
        
        \textbf{[31.2\%]} I got my kid vaccinated because vaccines are a win-win for me. & 
        \textbf{[21.0\%]} Yes, I will get the COVID-19 vaccine. (No, you cannot change my mind.) & 
        \textbf{[21.6\%]} my husband and I are 77 years old and have gotten both Covid vaccinations and we are just fine. \\
        \hline \hline

        \rowcolor[gray]{0.8}
        \multicolumn{3}{|c|}{\textbf{Topic 2 [Promote Vaccines]:} User is sharing positive emotions about vaccines.}\\
        \hline
        \rowcolor[gray]{0.9}
        \multicolumn{3}{|c|}{\textit{Top Words:} dose, go, receive, second, get, old, schedule, care, global, free}\\
        \hline

        \textbf{[26.8\%]} If you have certain health problems like heart or breathing conditions you can get a free flu vaccine. & 
        \textbf{[21.7\%]} Very good news on the Pfizer / BioNTech SE mRNA vaccine: early data shows 90\% effectiveness!! &
        \textbf{[25.2\%]} Starting April 5, Nevada is fast forwarding to opening up vaccinations to everyone 16+ \\
        \hline \hline

        \rowcolor[gray]{0.8}
        \multicolumn{3}{|c|}{\textbf{Topic 3 [Against Anti-Vaxxer]:} Showing disagreement with Anti-Vaxxers}\\
        \hline
        \rowcolor[gray]{0.9}
        \multicolumn{3}{|c|}{\textit{Top Words:} people, need, mask, risk, wait, pandemic, right, care, protect, save}\\
        \hline

        \textbf{[21.3\%]} fact - vaccine laws increase the vaccination rate.  fact - anti-vaxers believe in junk science. fact - your twitter bio made me lolz & 
        \textbf{[28.4\%]} Anti-vaxx conspiracy bollocks preys on marginalized people living in news deserts. I heard a person wants to take ivermectin - not COVID vaccine. & 
        \textbf{[30.4\%]} anti-vaxxers block access to the largest \#COVID vaccination site in LA shouting epithets at seniors waiting to get their \#CovidVaccine \\
        \hline \hline

        \rowcolor[gray]{0.8}
        \multicolumn{3}{|c|}{\textbf{Topic 4 [Supports Authorities]:} Thanking or supporting medical staff or frontline workers for vaccines.}\\
        \hline
        \rowcolor[gray]{0.9}
        \multicolumn{3}{|c|}{\textit{Top Words:} vaccineswork, help, leader, thank, polio, prevent, today, support, distribute, save}\\
        \hline

        \textbf{[20.7\%]} Thanks to funding and support from UNICEF, cold chain equipment, used to store vaccines & 
        \textbf{[28.9\%]} Thanks @user for being a champion in our mission to get COVID-19 tests \& vaccines to everyone & 
        \textbf{[22.8\%]} Just had my Pfizer jab !! good NHS thank you all and the Armed forces at the Vaccination point. \\
        \hline
    \end{tabular}
    \vspace{-2mm}
    \caption{Broad topics talked about by Pro-Vaxxers, along with the percentage of tweets in that topic among a time period (in brackets) and excerpts from tweets from the corresponding time periods.
    }
    \vspace{-2mm}
    \label{tab:annotation_themes_pro}
\end{table*}

\vspace{2mm}
\noindent \textbf{Topics discussed by Anti-Vaxxers:} The topics discussed by the Anti-Vaxxers are shown in Table~\ref{tab:annotation_themes_anti}, along with their distribution over the three time periods (considering the tweets that could be assigned topics), and example tweets.
We find that most of the broad topics have remained the same in the COVID times as they were in the pre-COVID times. 
However, the different topics are being discussed in different proportions in different time periods.

Across all three time-periods, many Anti-Vaxxers express \textbf{[Health Concerns]} about the vaccines, which is the most discussed topics across all the time periods. 
In the pre-COVID period, Anti-Vaxxers talked about how measles and flu vaccines cause various health concerns such as Autism and adverse effects on children. 
In the COVID times, the Anti-Vaxxers talk more about the ingredients of the COVID-19 vaccines which they assume to contain fetal cells, and how they will alter the DNA of those who take the vaccines.
Stemming from various concerns, they strongly oppose mandatory vaccination (the \textbf{[Against Mandatory]} topic in Table~\ref{tab:annotation_themes_anti}). 
The Anti-Vaxxers have been demanding freedom of choice of taking vaccines since the pre-COVID period. 
An emerging subtopic in 2021 is the protest against `vaccine passports' being required to visit various places.

Anti-vaxxers also frequently complain against the \textbf{[Big Pharma]} (large pharmaceutical companies) stating that they only care about profit (by pushing unnecessary vaccines), or do not not care about harmful effects of their vaccines. 
This criticism of Big Pharma seems to have reduced in the COVID-vax period, with people focusing more on other topics (discussed below).
A related topic is about the \textbf{[Political]} side of vaccines. This was talked about {\it slightly less in the Pre-COVID period}, with the discourse mostly limited to concerns that the Governments are not doing enough to prevent the big pharma from exploiting the people. 
However, in the COVID period, Anti-Vaxxers are a lot more vocal about the governments pushing vaccines to gain political support. 
Especially in the US, the 2020 presidential elections gave rise to a huge debate where the government was blamed for pushing COVID-19 vaccines (reflected by the increased fraction of such tweets in 2020).

Another common Anti-vax topic is that vaccines are \textbf{[Ineffective]} against the disease. 
In the pre-COVID period, uses opined that vaccines are ineffective against flu and measles.
In the context of COVID-19 vaccines, many Anti-Vaxxers users have expressed concern that these vaccines will be ineffective against the disease (e.g, since the virus can mutate). 
In the COVID-vax period even more users are concerned about the effectiveness, following reports of people affected by COVID even after getting vaccines.
Another major concern is that the vaccines are being rushed by the government and the big pharma, and that not enough testing has been done before releasing the vaccines to the general public (the topic \textbf{[Rushed]} in Table~\ref{tab:annotation_themes_anti}). 
Discussion on this topic has become much more frequent in the COVID times.

Finally, a few Anti-Vaxxers also talk about some more controversial topics. One of them is about \textbf{[Shedding]}, which says that vaccinated users transmit the disease to others. 
In pre-COVID period, this was a very frequent talking point against MMR and Flu vaccines.
In the COVID and COVID-vax periods, users also talk about shedding the virus after taking COVID-19 vaccines, but the frequency has reduced as compared to the pre-COVID times. 
Additionally, some \textbf{[Deeper Conspiracy]} theories are especially being discussed in the COVID times, such as COVID-19 being made up to trick people into buying vaccines, presence of tracking chips in vaccines, population control, etc. These conspiracy theories are talked about a lot more after the onset of COVID~\cite{cinelli2020covid}.

\vspace{1mm}
\noindent \textbf{Topics discussed by Pro-Vaxxers:} The topics discussed by Pro-Vaxxers are given in Table~\ref{tab:annotation_themes_pro}, along with their distribution across the three time-periods and some example tweets.
A lot of users have actively stated that they wish to take vaccines or have taken vaccines (\textbf{[Want Vaccines]}) and are seen to actively post information that promotes the adoption of vaccines (the \textbf{[Promote Vaccines]} topic in Table~\ref{tab:annotation_themes_pro}). 
In the COVID-vax period, many Pro-Vaxxers have shared that they have taken the COVID-19 vaccine, stating that the side effects they experienced are mild.
Pro-Vaxxers also \textbf{[Support Authorities]} by thanking them for development and distribution of vaccines. 
However, it is observed that the fraction of people actively talking about getting vaccines (the \textbf{[Want Vaccines]} topic) has reduced since the onset of COVID, which shows some level of vaccine hesitancy among the masses.
On the other hand, more Pro-vax people have actively expressed their disapproval against Anti-Vaxxers (the \textbf{[Against Anti-Vaxxer]} topic) in the COVID and COVID-vax periods (as also observed in other works~\cite{yuan2019examining,cossard2020falling}).

\vspace{1mm}
\noindent {\bf Summary:} We identified broad topics discussed by Anti-Vaxxers and Pro-Vaxxers across different time periods. 
Our identified topics match with those identified by prior works~\cite{mitra2016understanding,jamison2020not}. 
But to our knowledge, no prior work analyzed how Anti-vax / Pro-vax discussions have varied in pre-COVID times and COVID times.
While the broad topics being discussed are similar in pre-COVID and COVID times, we observe differences in their relative frequencies, e.g., lot more complaints about vaccines being rushed/ineffective and deeper conspiracy theories in COVID times, compared to the pre-COVID times.

\section{Change in vaccine-stances of users }
\label{sec:user_stances}

\begin{table}[t]
	\centering
\small
	\begin{tabular}{|l|c|c|c|}
		\hline
		& & \multicolumn{2}{c|}{\textbf{pre-COVID period (2018-19)}}\\
		\hline
		& & Anti-Vaxxers & Pro-Vaxxers \\
		\hline
		\multirow{2}{1.8cm}{\textbf{COVID period (2020)}} & Anti-Vaxxers & 2791 & \textbf{218} \\
		\cline{2-4}
		& Pro-Vaxxers & \textbf{21} & 6853\\
		
		\hline
		\hline
		
		& & \multicolumn{2}{c|}{\textbf{COVID period (2020)}}\\
		\hline
		& & Anti-Vaxxers & Pro-Vaxxers \\
		\hline
		\multirow{2}{1.8cm}{\textbf{COVID-vax period (2021)}} & Anti-Vaxxers & 7855 & \textbf{121}\\
		\cline{2-4}
		& Pro-Vaxxers & \textbf{329} & 9898 \\
		\hline
	\end{tabular}
	\vspace*{-2mm}
	\caption{Statistics of users who changed their vaccine-stance across the different time periods. The number of users who have changed their stance are marked in bold.}
	\label{tab:users_matrix}
	\vspace*{-4mm}
\end{table}

\begin{table*}[!t]
	\centering
	\footnotesize
	\begin{tabular}{|c|m{7.8cm}|m{7.8cm}|}
		\hline
		\rowcolor[gray]{0.8}
		 & \textbf{Pre-COVID period (2018-19):} Anti-Vaxxers &  \textbf{COVID period (2020):} Pro-Vaxxers\\
		\hline
		 U1 & 
		 Thank you, @USER. You know your son. Seizures followed vaccination. Vaccine safety has not been proven. Cases like yours demand further investigation. & 
		 My husband died after becoming ill with corona virus. I am glad a vaccine has been developed in record time. I intend to be vaccinated as soon as possible \\
		 \hline
		 U2 & 
		 Having an autistic child destroyed my family. I took my healthy baby to the doctor that gave the vaccine that caused it. &
		@USER Thank you for getting us vaccines in Illinois.  As a frontline healthcare worker I thank you. \\
		\hline \hline
		
		\rowcolor[gray]{0.8}
		 & \textbf{Pre-COVID period (2018-19):} Pro-Vaxxers & \textbf{COVID period (2020):} Anti-Vaxxers\\
		\hline
		U3 & 
		A vaccine eradicated smallpox. If the government was trying to kill us by way of a vaccine, we wouldn’t be living longer, and diseases would not be prevented &
		I will NEVER take a vaccine for COVID. This whole fiasco w/masks \& destroying our economy over a virus w/ 99\% survival rates, is ridiculous.\\
		\hline
		U4 & 
		Every child should be vaccinated I don’t understand why parents wouldn’t sad endangering others &
		Honestly this administration caused so much sickness and death I will not trust a vaccine.  \#LyingTrump \\
		\hline \hline
		
		\rowcolor[gray]{0.8}
		 & \textbf{COVID period (2020):} Anti-Vaxxers & \textbf{COVID-vax period (2021):} Pro-Vaxxers\\
		\hline
		U5 & 
		I'm not paying Big Pharma for a vaccine against something that they invented in order to sell vaccines. &
		I've had the first shot of the Phizer vaccine. I'm a school bus driver. I'm in contact with kids daily and several times\\ 
		\hline
		U6 & 
		But even a vaccine won't stop the spread, so we could see these levels for years. &
		Today has been a great day. My grandad has told he's getting the Covid vaccine. \\
		\hline \hline
		
		\rowcolor[gray]{0.8}
		 & \textbf{COVID period (2020):} Pro-Vaxxers & \textbf{COVID-vax period (2021):} Anti-Vaxxers\\
		\hline
		U7 & 
		Yes, it's bad, Covid is a nasty virus and I look forward to the vaccine rollouts. & 
		Covid is not dangerous to children. Forcing them to take this vaccine is insane.\\
		\hline
		U8 & 
		@USER Once we have a vaccine it will become as irrelevant as the common cold. Until then.... &
		They are of concern because some vaccines may be less than 50\% efficient in protecting from some variants.\\
		
		\hline
	\end{tabular}
	\vspace*{-2mm}	
	\caption{Samples of tweets (excerpts) from some stance-changed users (U1-U8). Each row shows two tweets posted by the {\it same user} during two different time periods. We do not state any usernames/identifiers to protect the privacy of users.}
	\vspace*{-2mm}
	\label{tab:stance_change_samples}
\end{table*}

Recall from Section~\ref{sub:user_classification} that we identified large numbers of Pro-Vaxxers and Anti-Vaxxers from the pre-COVID, COVID and COVID-vax  time periods (see Table~\ref{tab:users_identified}). 
Many of these users actively posted tweets only during one of the time periods; hence we could detect the stance of a user during {\it multiple time-periods} only for a smaller set of users.
We now consider those users for whom we could detect vaccine-related stance during at least two different time periods.
Table~\ref{tab:users_matrix} shows how their stances varied across the three periods.

Most users continue to follow the same stance towards vaccines. However, we find {\it 675 users who have changed their stance towards vaccines}. 
We see $218$ users who were pro-vaxxer in pre-COVID period (2018-19) have become Anti-Vaxxers in COVID period (2020), which corroborates with the rise of vaccine hesitancy in COVID times~\cite{bonnevie2020quantifying}. 
Interestingly, a small set of $21$ users did just the opposite -- they turned from Anti-Vaxxers in pre-COVID times to Pro-Vaxxers in 2020. 
Again, people seem to be gaining back trust in vaccines in 2021, as we see $329$ users shifting from Anti-Vaxxers (in 2020) to Pro-Vaxxers (in 2021). However, $121$ users who were supporting vaccines in 2020 started showing hesitancy towards vaccines in 2021. 
In fact, there are $14$ users who changed their stance multiple times -- from Pro-Vaxxers in 2018-19 to Anti-Vaxxers in 2020 and again to Pro-Vaxxers in 2021.

\vspace{1mm}
\noindent \textbf{Checking for bots:}
\reviewerchange{
Automated bots abound on Twitter~\cite{sayyadiharikandeh2020detection},
and we wanted to ensure that we are focusing on human users. 
To this end, we used Botometer-v4~\cite{sayyadiharikandeh2020detection}  which returns a score between $0$ (very likely to be a human) and $1$ (very likely to be a bot) for a Twitter user.
We applied Botometer-v4 over all the $675$ users who changed their stance.
The tool assigned a score $\geq 0.5$ to $68$ users, implying that these users are more likely to be bots. 
Since the automated detection of bots often predicts a lot of false positives~\cite{rauchfleisch2020false},
we manually checked these $68$ accounts. 
We found only {\it one} of these accounts to be automated, that automatically shares articles from an external personal blog. 
We ignore this account in subsequent analyses.
}

\vspace{1mm}
\noindent In the rest of this section, we focus on the $674$ users who have changed their vaccine-related stance. We refer to this set of users as the \textit{\textbf{stance-changed users}}.

\begin{table}[t]
	\centering
	\footnotesize
	\begin{tabular}{|c|m{76.5mm}|m{76.5mm}|}
		\hline
		 & \textbf{Against a Vaccine} &  \textbf{Supporting a vaccine}\\
		\hline
		\hline
		 U1 & 
		 declining \textbf{Pfizer} vaccine after seeing two peoples side effects &
		 I’m more likely to trust the \textbf{Oxford} one \\
		 \hline
		 U2 & 
		 I do have concerns about the efficacy of the \textbf{AstraZeneca} vaccine though, if it gets rolled out. &
		 There's a shortage of \textbf{Pfizer} vaccines and the \textbf{Moderna} vaccine won't be rolled out till March \\
		 \hline
		 U3 & 
		 I don’t want the \textbf{Pfizer} vaccine. Let them have it. &
		 PLEASE use the \textbf{Moderna} vaccine, not the \textbf{Pfizer} or others \\
		 \hline
		 U4 & 
		 {\bf Pfizer} has not been tested, at all, at these intervals. &
		 {\bf Oxford} vaccine already demonstrating reduced transmission \\
		\hline
	\end{tabular}
	\vspace{-2mm}
	\caption{Samples of tweets (excerpts) from some of the stance-changed users who posted against a particular vaccine, and supporting another vaccine. 
	}
	\label{tab:different_vaccines}
	\vspace{-4mm}
\end{table}

\vspace{2mm}
\noindent \textbf{Is the stance change an effect of users posting different opinions about different vaccines?}
\new{
It is possible that an individual user has different opinions about different COVID-19 vaccines.
Hence there is a potential concern about the set of stance-changed users -- is the observed change in stance an artefact of users posting positive opinions about one vaccine at one point of time, and negative opinions about some other vaccine at some other point of time?}

To address this concern, we check for the presence of different vaccine names (along with different variations of their names and manufacturers, such as AstraZeneca/Oxford, Pfizer/BioNTech, Moderna, Sputnik, Sinopharm, etc.) in all the tweets posted by the stance-changed users.
Out of the $674$ distinct stance-changed users, $133$ had posted at least one tweet about two different vaccines. 
Out of these, we found only $13$ users (2\% of $674$) who posted a clear preference towards one vaccine, while rejecting/opposing another vaccine. 
We have given a few examples of tweets from such users in Table~\ref{tab:different_vaccines} (each row showing two tweets from the same user). 
We also noted that, all these users also expressed concerns about vaccines in general during one of the time periods, and posted in support of vaccines in general in some other time period.
Thus, we conclude that all the $674$ users whom we have identified, actually changed their \textit{general} (not related to specific vaccines) stance towards vaccines at some point of time.

\begin{figure}[t]
	\centering
	\begin{subfigure}[b]{0.35\linewidth}
		\centering
		\includegraphics[width=\textwidth,height=5cm]{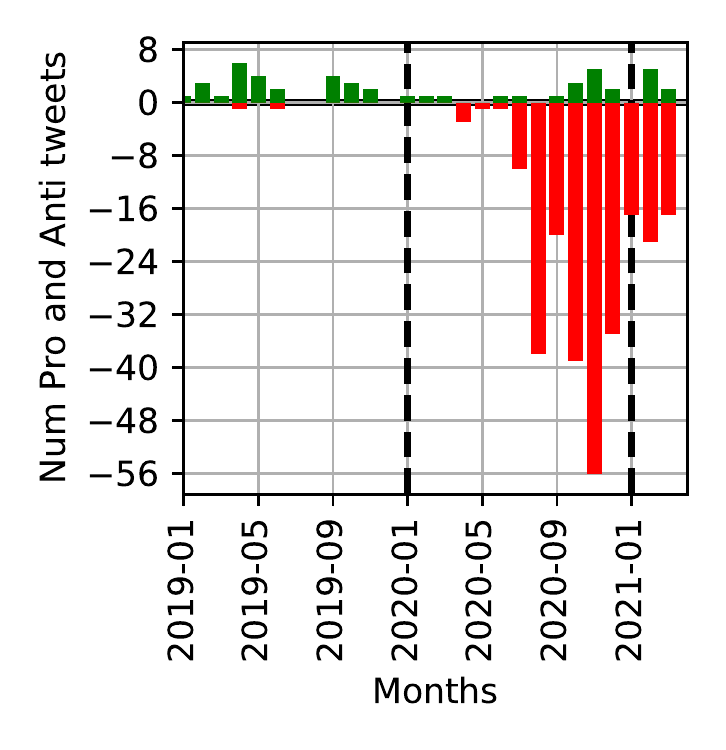}
		\caption{Changed from pro (2018-19) to anti (2020)}
		\label{fig:u_pa1}
	\end{subfigure}
	\begin{subfigure}[b]{0.35\linewidth}
		\centering
		\includegraphics[width=\textwidth,height=5cm]{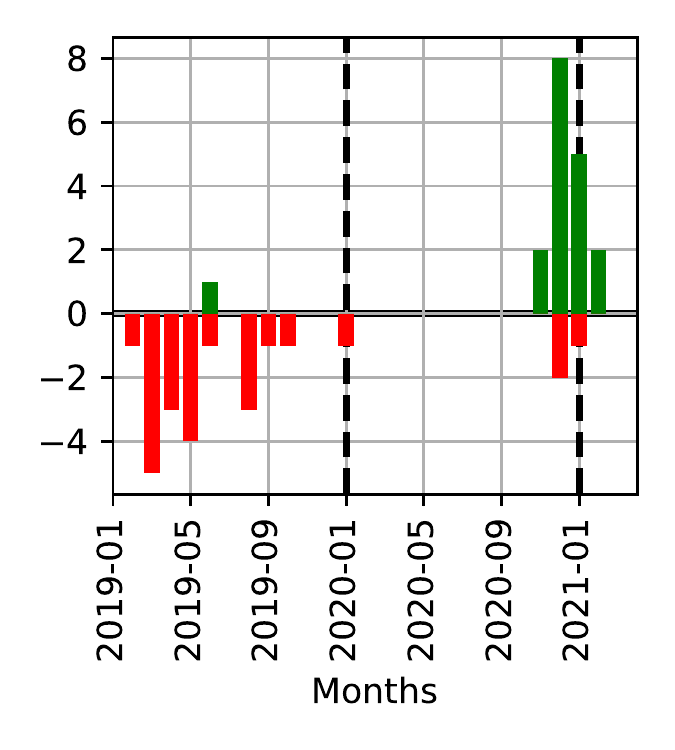}
		\caption{Changed from anti (2018-19) to pro (2020)}
		\label{fig:u_ap1}
	\end{subfigure}
    \vspace*{-2mm}
	\caption{Number of anti-vax (-ve, in red) and pro-vax (+ve, in green) tweets posted by a few stance-changed users. The dashed vertical lines represent the start of the time periods -- COVID (2020) and COVID-vax (2021).}
	\label{fig:stance_change_samples}
	\vspace*{-2mm}
\end{figure}

\vspace{2mm}
\noindent \textbf{Visualising the change for some individual users:}
Figure~\ref{fig:stance_change_samples} shows the number of pro-vax and anti-vax tweets posted per month by some of the stance-changed users. The X-axis shows the month, and the Y-axis shows the number of pro-vax tweets (green bars on the positive side) and anti-vax tweets (red bars on the negative side) posted by the user.

Table~\ref{tab:stance_change_samples} shows samples of tweets posted by some of the stance-changed users. Each row of this table show two tweets posted by the same user during two different time periods.
Some of the tweets seem to indicate potential reasons of why the users changed their stances. 
For instance, user U1 (who shifted from Anti-Vaxxers in pre-COVID period to Pro-Vaxxers in COVID period) experienced a life-changing event that may have changed his/her stance. 
Those who changed from Pro-Vaxxers to Anti-Vaxxers (user U3 \& U4) during this time 
were getting skeptical of COVID-19 vaccines in specific due to various reasons, including political factors. 
Some users were hesitant about the COVID-19 vaccines in 2020; but with the actual rollout of vaccines in 2021, they are willing to get vaccinated (e.g., U5 \& U6).
Finally some users who were Pro-Vaxxers in 2020 also have turned Anti-Vaxxers in the COVID-vax period (U7 \& U8), due to factors such as mistrust in how effective vaccines would be, and opposition to mandatory vaccination.

\subsection{Representativeness of stance-changed users}

\begin{figure*}[t]
	\centering
	\begin{subfigure}[b]{0.35\textwidth}
		\centering
		\includegraphics[width=\textwidth]{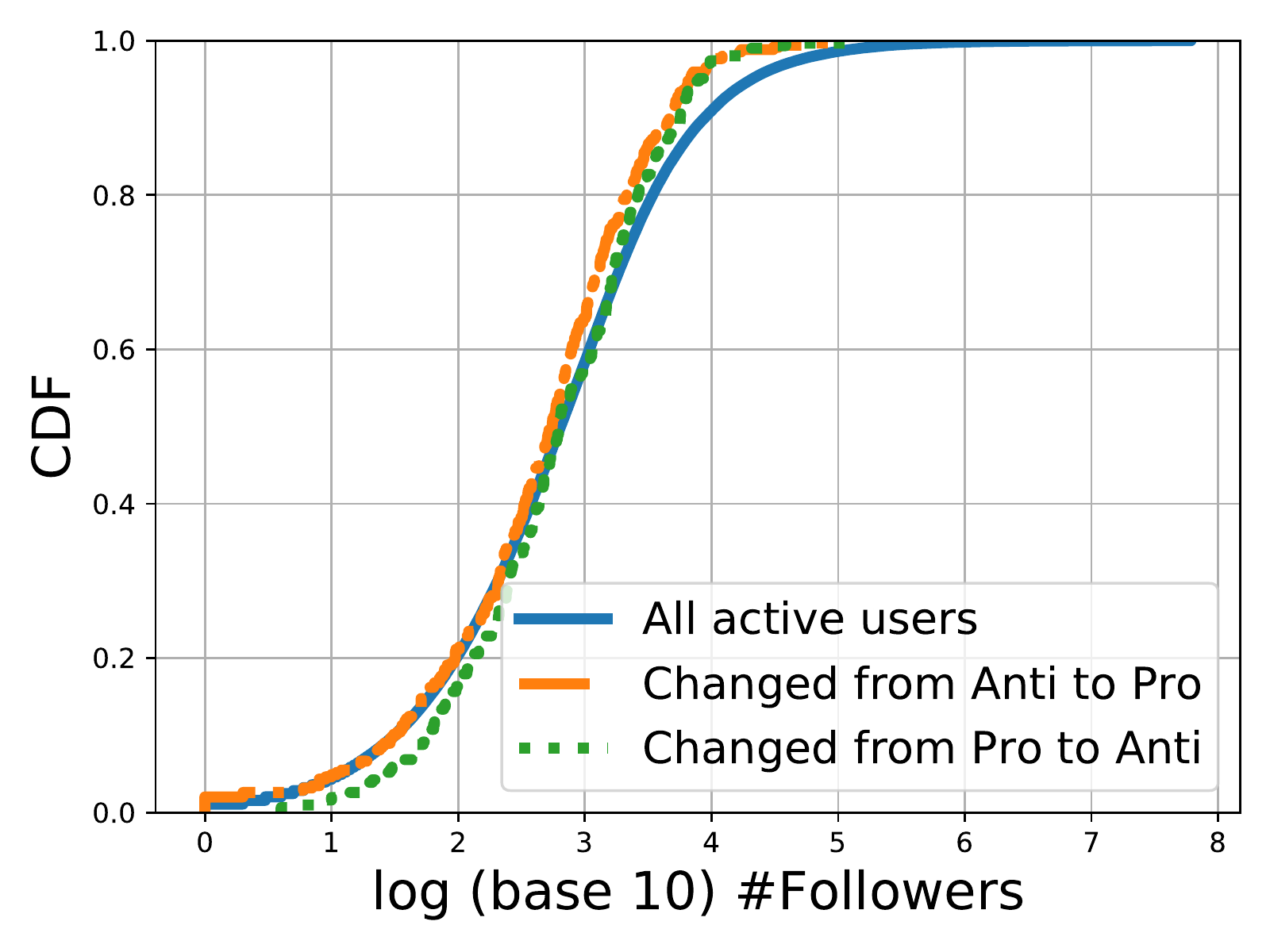}
		\caption{CDF of log \#Followers}
		\label{fig:cdf_follower}
	\end{subfigure}
	\begin{subfigure}[b]{0.35\textwidth}
		\centering
		\includegraphics[width=\textwidth]{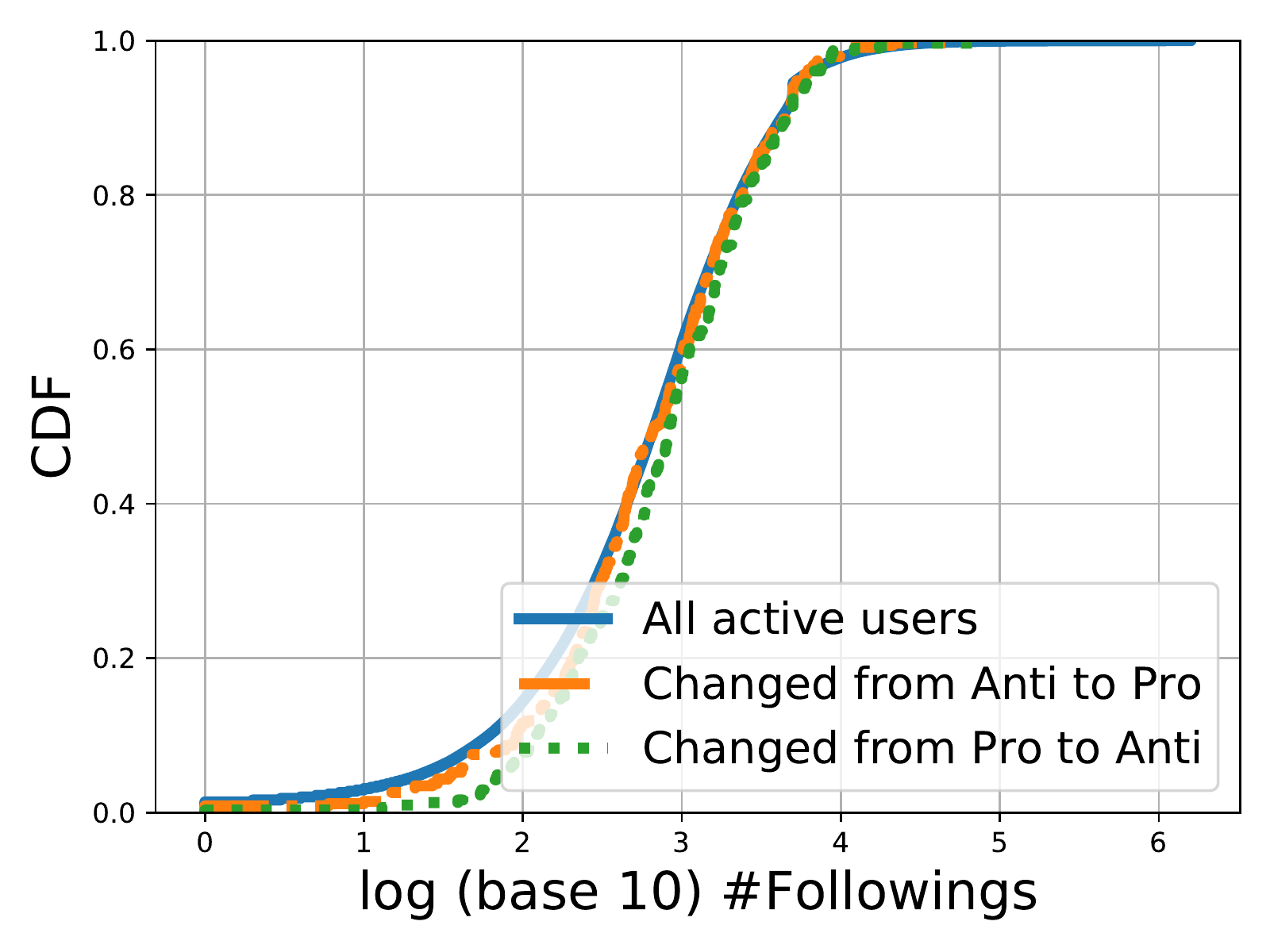}
		\caption{CDF of log \#Followings}
		\label{fig:cdf_following}
	\end{subfigure}
	\begin{subfigure}[b]{0.35\textwidth}
		\centering
		\includegraphics[width=\textwidth]{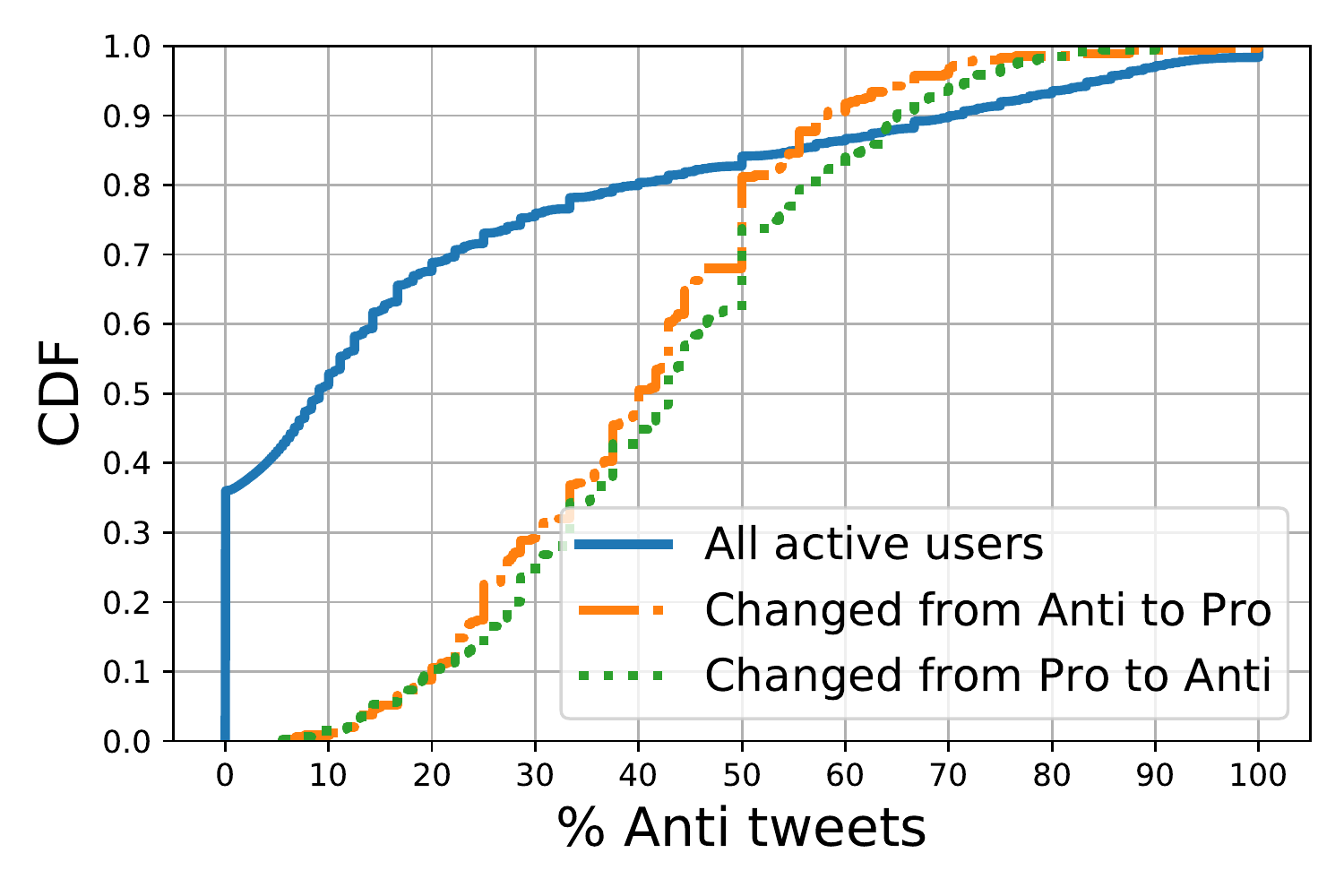}
		\caption{CDF of \%Anti Tweets posted}
		\label{fig:cdf_anti}
	\end{subfigure}
	\begin{subfigure}[b]{0.35\textwidth}
		\centering
		\includegraphics[width=\textwidth]{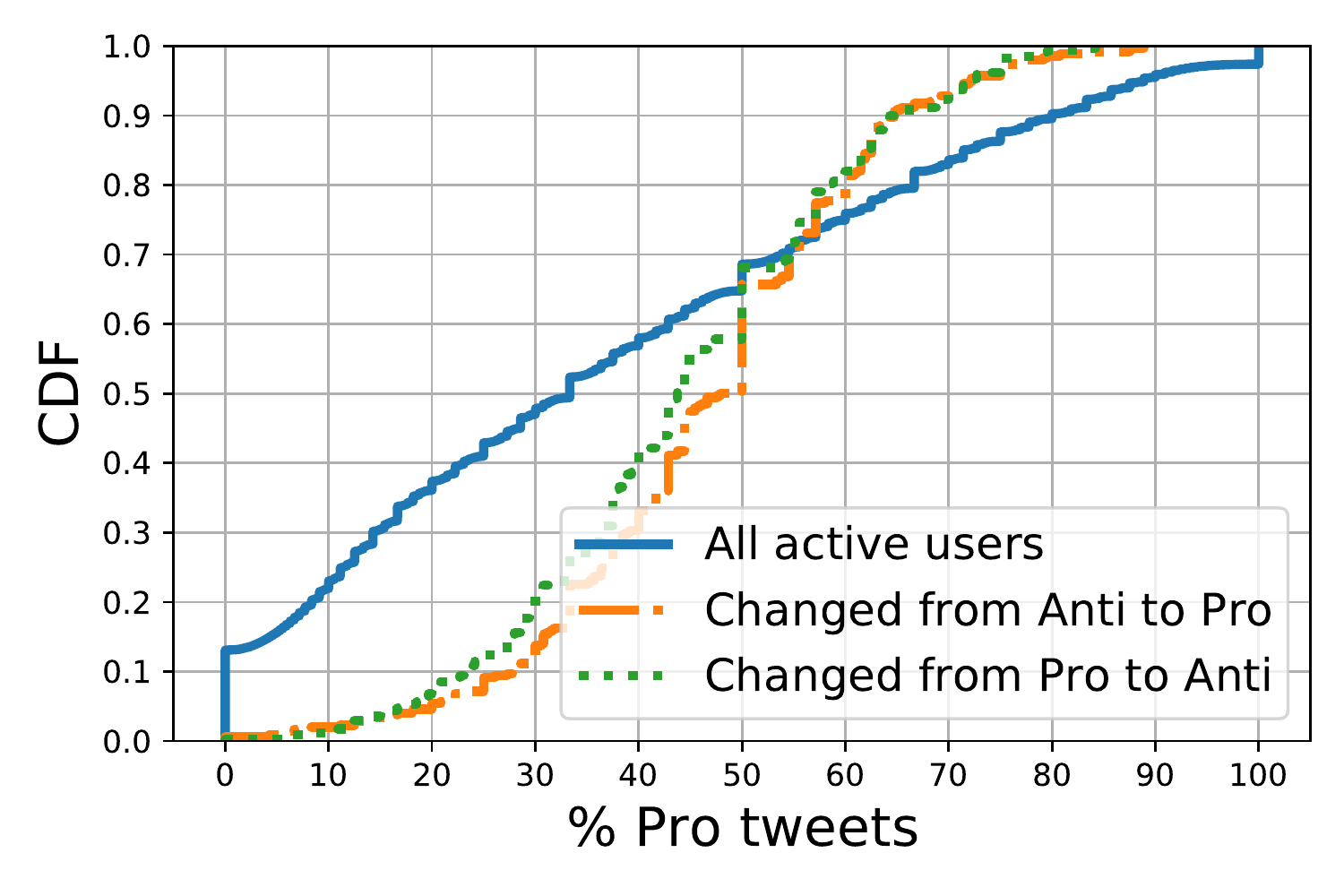}
		\caption{CDF of \%Pro Tweets posted}
		\label{fig:cdf_pro}
	\end{subfigure}
	\vspace*{-2mm}
	\caption{\new{Comparison of stance-changed users with the general active Twitter population interested in vaccines, in terms of (a)~number of followers, (b)~number of followings, (c)~\% of Anti-vax tweets posted, and (d)~\% of Pro-vax tweets posted.}}
	\label{fig:user_representativeness}
	\vspace*{-4mm}
\end{figure*}

\new{In this section, we analyse the stance-changed users to understand {\it how representative they are of the general population of active Twitter users}. 
This question is important to ascertain the  practical utility of the insights about these users (as reported later in the paper), for policy-makers.}

As stated in Section~\ref{sub:user_classification}, we attempted to detect the stance of a user during a certain time-period only if that user posted at least 3 tweets during that period. 
Thus the stance-changed users have posted at least 3 tweets each in two different time periods.
Hence, we compare these users with the population of active Twitter users who posted at least 6 tweets related to vaccines in the entire time period of analysis (Jan 2018 -- March 2021). 
In total, there are $~340K$ such active users.

Figure~\ref{fig:user_representativeness} compares social network properties (number of followers and followings) and the tweet composition behaviour of the two user-groups.
Figures~\ref{fig:cdf_follower} and \ref{fig:cdf_following} show that the followers and followings distributions of the stance-changed users and the active users are very similar.

Next, we compare the tweeting behavior of the two user-groups; Figure~\ref{fig:cdf_anti} and Figure~\ref{fig:cdf_pro} respectively compare the distributions of the fraction of Anti-Vax tweets and fraction of Pro-Vax tweets posted by an individual user (in the entire time period of analysis) in the two user-groups. 
We see that a large portion of the active user population (37\%) did not post any Anti-Vax tweet at all (Figure~\ref{fig:cdf_anti}), while about 12\% of the active user population did not post any Pro-Vax tweet at all (Figure~\ref{fig:cdf_pro}); also, only a small fraction of the active users post a large fraction of Anti-Vax tweets.
In sharp contrast, the stance-changed users post higher fractions of Anti-vax as well as higher fractions of Pro-vax tweets, compared to the general active Twitter population interested in vaccines (who post a lot more neutral tweets on vaccines).

Thus, the stance-changed users are representative enough of the active Twitter population interested in vaccines in terms of social network properties, but they express their opinions (both Anti-vax and Pro-vax) much more than the general active Twitter population who post about vaccines.

\begin{figure}
    \centering
    \includegraphics[width=0.7\linewidth, height=7cm]{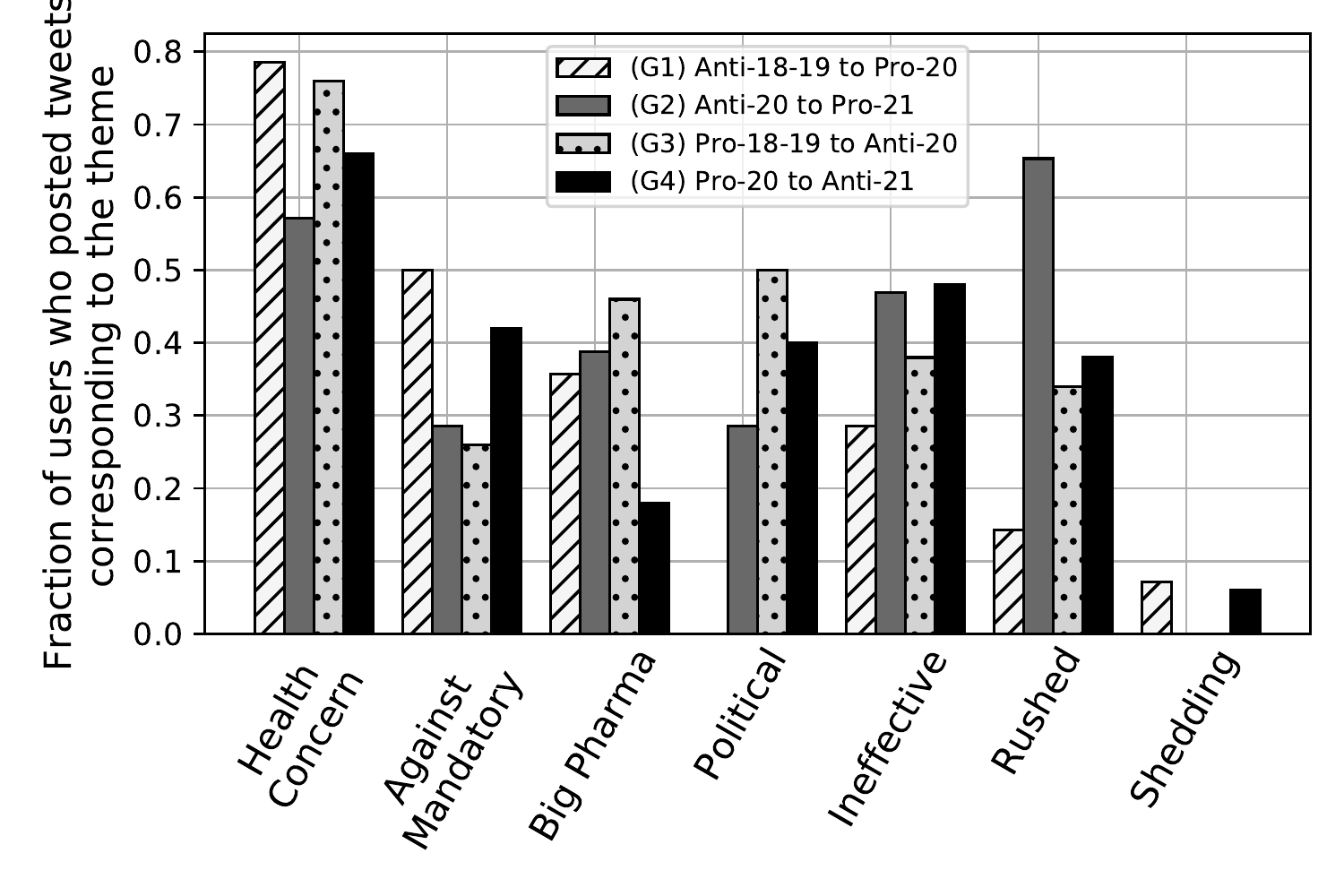}
    \vspace*{-6mm}
    \caption{\new{Anti-vax topics posted by stance-changed users}}
    \label{fig:themes_com}
    \vspace*{-4mm}
\end{figure}

\subsection{Characterizing stance-changed users}
Now, we analyze the stance-changed users to understand what anti-vax topics they used to discuss before/after their stance change.
To this end, we select four groups of stance-changed users -- 
(G1)~all the $21$ users who changed from Anti-Vaxxer in pre-COVID period to Pro-Vaxxer in COVID period, 
(G2)~$50$ randomly selected users from those who changed from Anti-Vaxxer in COVID period to Pro-Vaxxer in the COVID-vax period, 
(G3)~$50$ random users from those who changed from Pro-Vaxxers to Anti-Vaxxers between the pre-COVID and COVID periods, and 
(G4)~$50$ random users who changed from Pro-Vaxxers in COVID prioed to Pro-Vaxxers in COVID-vax period. 
Two authors manually examined the vaccine-related tweets posted by these users from the time period when they were Anti-Vaxxers, to identify which anti-vax topics (out of the ones we identified in Table~\ref{tab:annotation_themes_anti}) they posted about.

Figure~\ref{fig:themes_com} compares these groups of users -- the anti-vax topics are shown on the X-axis while the fraction of users from each group who posted tweets corresponding to the topics is shown on the Y-axis.
Note that we did {\it not} see any tweet about deeper conspiracy theories (Topic~8 in Table~\ref{tab:annotation_themes_anti}) from any of these groups. Since all these users were/became Pro-Vaxxers at some point, it is likely that they were {\it not} staunch Anti-Vaxxers, and conspiracy theories  may be an opinion prevalent among staunch Anti-Vaxxers only.

\vspace{2mm}
\noindent \textbf{Users who shifted from Anti- to Pro-Vaxxers:}
We observe an interesting difference in the anti-vax opinions of the two groups of users who changed from Anti- to Pro-Vaxxers (G1 and G2).
Most of the users in G1 -- who were Anti-Vaxxers in pre-COVID period (2018-19) -- opposed vaccines because of the health concerns, mostly in regards to their children being harmed. 
A fair number among them also were against mandatory vaccination, and against how the Big Pharma are trying to sell vaccines with side effects. 
However, the users in G1 did {\it not} post much about the vaccine development being rushed/untested or the role of politics.
On the contrary, users in G2 -- Anti-Vaxxers in COVID period (2020) but supporting COVID-19 vaccines (in 2021) -- seem to care a lot more about the political nature of the vaccines, the potential ineffectiveness of the vaccines against COVID, and that the vaccine development process was rushed.

\vspace{2mm}
\noindent \textbf{Users who shifted from Pro- to Anti-Vaxxers:}
For the users in G3 -- who became Anti-Vaxxers in the COVID period -- 
the main concerns seem to be about the potential side-effects of vaccines (health concerns), and that the big pharma and politicians are exploiting the crisis to their advantage.
Their other primary concerns are about the rushing of vaccines into distribution, and that the vaccines would be ineffective against the disease.
On the other hand, among the users in G4 -- who shifted from Pro-vaxxer in 2020 to Anti-Vaxxers in the COVID-vax period -- most blamed the governments (more than the big pharma), stating that they will not take vaccines offered by specific governments and political leaders. 
When vaccinations actually started, these users started to tweet more against mandatory vaccination, and also being skeptical about the effectiveness of the vaccines due to rising mutations of the corona-virus. Health concerns about vaccines remains a common theme between both user-groups.

\vspace{2mm}
\noindent
\new{
\textbf{Users who were Anti-Vaxxers in 2020 but Pro-Vaxxers in other time periods:}
Finally, we compare the two groups G2 and G3 -- both these groups were Anti-vaxxers in 2020; however, users in G2 became Pro-vaxxers in 2021, whereas users in G3 were Pro-vaxxers in 2018-19. 
Users in G3 tweeted more in 2020 about Health concerns and the Political side of vaccines, than the users in G2. 
In contrast, the users in G2 tweeted (in 2020) more about vaccines being Ineffective and Rushed. 
These observations signify that even though both groups had the same stance (Anti-Vax) in the COVID period (2020), their concerns were different.
}

\subsection{Exploring correlation of users' stance with stance of neighbours}

We explore if the social network has an effect of the changing stance of users. In particular we analyse the influence (if any) of the people whom a user is following at various points of time, on the changing stance of the said user. 

\vspace{2mm}
\noindent {\bf Method:} Currently, it is not possible on Twitter to know the creation date for social links, or to collect the list of followings of a user (also known as  \textit{friends} of the user) on a past date. Hence, we adopt the following method. 
For a particular user $u$, we collect all the current followings of $u$. Then we check the tweets posted by each of these followings during a particular time period (e.g., 2018-19, or 2020, or 2021). We compute what fraction of $u$'s followings were Pro-Vaxxers and Anti-Vaxxers during each of the three time periods, based on the vaccine-related tweets that they (the followings) posted in the respective time periods.

Thus, for user $u$, we first compute the fractions of Anti-Vaxxer and Pro-Vaxxer followings of $u$ during a particular time period. 
Then we see how the fractions of Anti-Vaxxer and Pro-Vaxxer followings of $u$ have {\it changed} from one time period to another (e.g., from 2018-19 to 2020, or from 2020 to 2021).
Finally, we average this change in fractions of Anti-Vaxxer and Pro-Vaxxer followings of a user, 
over all users in a group (e.g., the groups G1, G2, G3, G4 introduced in the previous section).

\begin{table}[t]
	\centering
    \small
	\begin{tabular}{|l|cc|cc|}
		\hline
		 User-groups  & \multicolumn{4}{c|}{Change in \#followings who were} \\
		 \cline{2-5}
		 who changed & \multicolumn{2}{c|}{2018-19 to 2020} & \multicolumn{2}{c|}{2020 to 2021}\\
		 \cline{2-5}
		 their stance & Pro  & Anti & Pro & Anti\\
		\hline
        G1: Anti (2018-19)  & 1.06$\times$ & 1.94$\times$ & 1.45$\times$ & 0.70$\times$ \\
        to Pro (2020) & (0.82, 1.29) & (1.03, 2.85) & (0.99, 1.91) & (0.47, 0.93) \\
        \hline
        G3: Pro (2018-19) & 1.04$\times$ & 4.43$\times$ & 1.67$\times$ & 0.85$\times$ \\
        to Anti (2020) & (0.88, 1.20) & (3.60, 5.27)& (1.47, 1.86)& (0.75, 0.94)\\
        \hline \hline
        G2: Anti (2020) & 1.34$\times$ & 4.38$\times$ & 2.25$\times$ & 0.70$\times$ \\
         to Pro (2021) & (1.20, 1.48) & (3.43, 5.34) & (2.10, 2.40) & (0.60, 0.79) \\
        \hline
        G4: Pro (2020) & 1.38$\times$ & 5.89$\times$ & 1.49$\times$ & 1.25$\times$ \\
        to Anti (2021) & (1.19, 1.56) & (4.01, 7.78) & (1.30, 1.67) & (1.06, 1.45) \\
        \hline
        \hline
        Pro (2018-19) to Anti & 0.72$\times$ & 1.94$\times$ & 2.17$\times$ & 0.35$\times$ \\
        (2020) to Pro (2021) & (0.56, 0.87) & (1.16, 2.73) & (1.61, 2.74) & (0.11, 0.58) \\
        \hline
	\end{tabular}
	\vspace{-2mm}
	\caption{\new{Studying the change of stance among the followings of the stance-changed users -- shown are the changes in the number of anti-vax and pro-vax followings between different time periods, averaged over all users in a particular group. 95\% confidence intervals are given below each number in parantheses.}}
	\vspace{-4mm}
	\label{tab:friend_pop}
\end{table}

\vspace{2mm}
\noindent {\bf Observations:}
\new{
Table~\ref{tab:friend_pop} shows these changes. 
For instance, across all the users in G1 (who changed from Anti-Vaxxer in 2018-19 to Pro-Vaxxer in 2020), the fraction of Pro-Vaxxers in their followings increased in 2020 to $1.06$ times of the fraction in 2018-19, whereas the fraction of Anti-Vaxxer followings increased to $1.94$ times over the same time-period.
Below each number, we have also added the 95\% confidence intervals (within parentheses) to demonstrate that the reported numbers are meaningful, and not skewed because of a few particular users.
}
The first notable observation from Table~\ref{tab:friend_pop} is that only a few of the changes are close to $1.0$. This implies that, for the stance-changed users, the amount of Anti-Vaxxers and Pro-Vaxxers among their followings changed by quite a margin across different time periods.

In most cases, we also see correlation between the change of stance of the users with the change in the fraction of followings of the corresponding stance. 
Even though the overall number of Anti-Vaxxers increased in 2020, and Pro-Vaxxers in 2021 (as was seen in Section~\ref{sub:user_classification}), we see a difference in the amount of increase in the fractions between the different user groups. 
For example, among the users in G3 (who shifted from Pro-Vax in 2018-19 to Anti-Vax in 2020), the fraction of Anti-Vaxxers in their followings increased by $4.43$ times in 2020. 
In contrast, for the users in G1 (who shifted from Anti-Vax in 2018-19 to Pro-Vax in 2020), the increase of Anti-Vaxxers in their followings was only $1.94$ times.
Similarly, for the users in G4 (who changed from Pro-Vax in 2020 to Anti-Vax in 2021) and vice versa (G2), the increase in Pro-Vaxxers in the followings (from 2020 to 2021) was $1.49$ times and $2.25$ times respectively. 
The most interesting observation is for those users who changed their stance from Pro-Vaxxers to Anti-Vaxxers and again back to Pro-Vaxxers -- there is an increase of $1.94$ times in the  fraction of Anti-Vaxxers among their followings between the pre-COVID and COVID periods, whereas the fraction of Pro-Vaxxers among their followings increased by $2.17$ times between the COVID and the COVID-vax periods.

We also analysed the users who were Pro-Vaxxers or Anti-Vaxxers {\it throughout the three time periods} (details omitted for brevity). Even though the amount of Pro-Vaxxers and Anti-Vaxxers in their followings change over the time periods, the fraction of followings with the {\it same stance as the user} hugely surpasses the fraction of followings with the opposing stance throughout all time periods. For the users who are Pro-Vaxxers throughout, the fraction of Pro-Vaxxer followings was $80$ times that of the fraction of Anti-Vaxxer followings at minimum (COVID period), and $230$ times at maximum (pre-COVID period). 
Similarly, for the users who are Anti-Vaxxers throughout, the fraction of Anti-Vaxxer followings was $4$ times the fraction of Pro-Vaxxer followings at minimum (pre-COVID period) and $7$ times at maximum (COVID-period).

\new{
Even though the overall number of users posting about vaccines may have increased during COVID times, the important point to note is that in most cases, there is a much greater increase in the fraction of followings with the same stance as the user, than the increase in followings with the opposite stance.
}
Thus, the vaccine-stance of a user seems to be correlated with the vaccine-stance of his/her social neighborhood. Also, our findings hint that users who change their vaccine-stances have a non-homogeneous neighborhood across pre-COVID and COVID times.

\section{Conclusion}
\label{sec:conclusion}

This work presents a first attempt to systematically analyse 
how a global event (the COVID-19 pandemic) has
changed the vaccine-related stances/opinions of people from pre-COVID to COVID times.
Our findings also hint at how the stance-change of a user can be correlated with factors such as major events in his/her life (e.g., death of a close relative) or changes in his/her social neighborhood.

\vspace{2mm}
\noindent \textbf{Implications of the work:}

We believe that the present work is useful to the authorities in their quest to ensure large-scale adoption of vaccines, by educating and disseminating specific information to specific people in the society. 
Essentially, we provide a framework to use social media data to automatically detect the vaccine-stance of users at scale, and understand the primary vaccine-related concerns (if any) of users in different time periods. 
This framework can help the authorities in the following ways --
(i)~{\bf Devising targeted approaches to alleviate people's vaccine-related concerns:} A policy maker cannot educate people unless they know the reasons/concerns behind their vaccine-related stance. 
We have seen that the anti-vax concerns are very different, from deep-rooted (e.g., vaccines kill people) to contextual (e.g., concerns regarding  COVID-19 vaccines being ineffective and rushed).
So, the policy makers need a targeted approach to educate people having different types of concerns. 
(ii)~{\bf Identifying and reaching out to people who changed their stances:} We showed there is a population who have changed their vaccine-stances. The authorities can identify the people who can potentially move over from anti-vax to pro-vax, and better nudge those people towards vaccination.
Perhaps, the experiences of some of the users who changed from Anti-vaxxer to Pro-vaxxer, can be used to nudge others in the Anti-vax community towards vaccination. 
Also, the authorities can investigate why some originally pro-vax users have become hesitant to take COVID-19 vaccines, and which specific concerns need to be clarified for them.
(iii)~{\bf Utilizing social media for nudging people towards vaccination:} 
We found that the social connections can potentially influence the stance of users, which suggests the importance of using popular social media users in promoting more Pro-Vax information. However, the information needs to be carefully designed in accordance to the primary concerns of different target user groups, without which they might just be mistaken as propaganda. 
Also, policy makers can design targeted friend/content recommendation algorithms to break into today's Anti-vax echo chambers with pro-vax opinions.

\vspace{2mm}
\noindent
\textbf{Future work:} We only explored the effects of the stances of a user's followings on the stance change of the user. 
Other forms of interactions such as retweeted or replied to/by can also have an effect on the stance change of users, and can be explored as  a future work.
Another future work would be to design predictive models that attempt to predict which users are likely to change their stance, e.g., from a user's posts and the stances of the followings of a user.

\vspace{4mm}
\noindent {\bf Acknowledgements:}
The authors acknowledge the anonymous reviewers whose comments greatly helped to
improve the paper. 
The project is partially supported by a research grant from Accenture Corporation.
The first author (S. Poddar) is also supported by the Prime Minister's Research Fellowship (PMRF) from the Ministry of Education, Government of India.

\bibliographystyle{apalike}
\bibliography{ref}

\end{document}